\documentclass[twocolumn, prl, aps, floats]{revtex4}
\usepackage{amsmath,amssymb,graphicx,nicefrac,tabularx,xcolor}
\def\A{\mathbf{A}}
\def\At{\tilde{A}}

\def\qt{\tilde{q}}
\def\Qt{\tilde{Q}}

\begin{document}

\title{Deep-layered machines have a built-in Occam's razor}
\author{Thomas Fink}
\affiliation{London Institute for Mathematical Sciences, Royal Institution, 21 Albemarle St, London W1S 4BS, UK}

\begin{abstract}
\noindent
Many complex systems map vast numbers of microscopic descriptions onto far fewer macroscopic outcomes.
Surprisingly, these input-output maps strongly favor simple outputs.
In an exactly solvable deep-layered machine, we prove that simplicity bias strengthens with every layer. 
The probability of an output function $F$ approaches $\Pr(F)\propto 2^{-K(F)}$,
where $K$ is the Kolmogorov complexity.
Our results show that hierarchical composition alone can generate an intrinsic Occam's razor, 
suggesting a general mechanism for simplicity bias in learning and evolution.
\end{abstract}

\maketitle
\begin{figure}[b!]
	\centering
	\includegraphics[width=0.99\columnwidth]{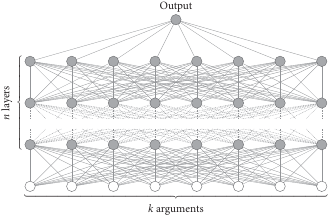}
	\caption{\small
	\textbf{Deep-layered machine.}
	In a deep-layered machine of width $k$ and depth $n$,
	each of the nodes (gray circles) computes a random Boolean function of the $k$ nodes below it.
	We prove that, as $n$ increases, the distribution of output functions becomes increasingly biased 
	toward functions with low Kolmogorov complexity $K$.
	At depths of order $2^k$, the probability of an individual output $F$ approaches $\Pr(F)\propto 2^{-K(F)}$, 
	before collapsing onto the constant functions true and false.
	}
	\label{Architecture}
\end{figure}
\begin{figure}[b!]
	\includegraphics[width= 0.99\columnwidth]{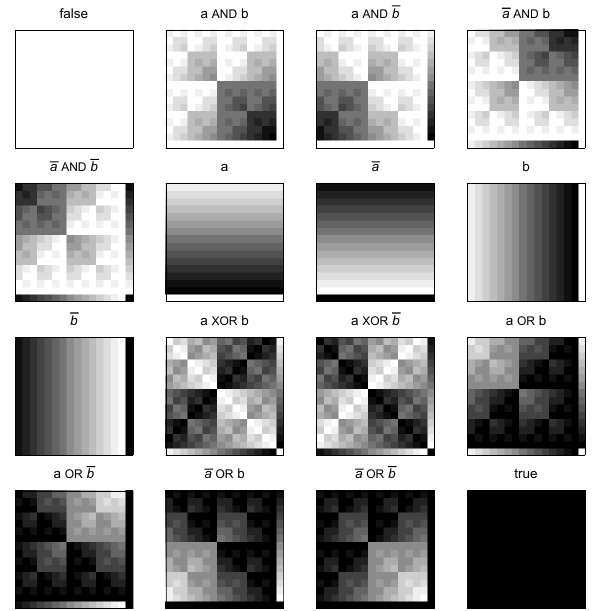}
	\caption{\small
	\textbf{Input-output map for two arguments.}
	For $k=2$ arguments and depth $n=1$, 
	our machine maps $16^3$ configurations of local Boolean functions onto just 16 possible output functions, 
	represented by different shades of gray.
	Each panel corresponds to one of the 16 Boolean functions at the top node, 
	while its $16\times16$ squares enumerate the choices for the two nodes below.
	The relative frequencies of the shades give the probability distribution of the output.
	}
	\label{HeroShot}
\end{figure}
\begin{figure}[b!]
	\includegraphics[width= 1\columnwidth]{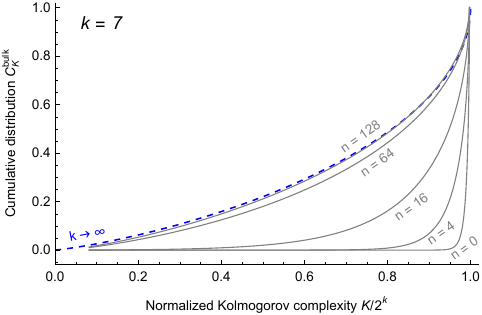}
	\caption{
	\textbf{Cumulative complexity distribution.}
	For width $k = 7$ and increasing depths $n$, 
	we plot the cumulative distribution $C_K^{\mathrm{bulk}}(n)$ that 
	a nonconstant output has Kolmogorov complexity at most $K$, against $K/\ell$.
    	The distributions shift towards lower complexity, 
	approaching the large-$k$ asymptotic distribution $2H_2^{-1}(K/2^k)$ (blue dashed curve) 
	at depths of order $2^k$.
    	We approximate the Kolmogorov complexity by $K=\log_2\binom{2^k}{w}+k/2$, 
	where $w$ is the Hamming weight and $k/2$ is a finite-$k$ correction.
}
	\label{ComplexityDistribution}
\end{figure}

\noindent 
Input-output maps are prevalent in biology, physics, mathematics and technology. 
In an input-output map, a large space of microscopic descriptions generates a much smaller space of macroscopic outcomes.
Examples range from 
biological evolution, where many genotypes map onto the same phenotype \cite{Li1996, Greenbury2022}, 
to neural networks, where many parameter configurations implement the same output function \cite{Mozeika2020, VallePerez2029}.
\\ \indent
A striking empirical observation is that input-output maps are not neutral;
rather, random inputs are exponentially more likely to generate simple outputs than complex ones.
Here, simplicity refers to low Kolmogorov complexity, that is, a short algorithmic description.
This intrinsic Occam's razor has been observed in computational and biological systems \cite{Dingle2018},
genotype-phenotype maps \cite{Johnston2022} and
deep neural networks \cite{Mingard2025}.
\\ \indent
Despite its prevalence, to date simplicity bias has been observed only empirically. 
We still lack a solvable model that explains why random microscopic rules preferentially generate simple macroscopic behaviors.
\\ \indent
To isolate the mechanism underlying simplicity bias, we study a model
that we can solve exactly: a deep-layered machine in which each node
computes a random Boolean function of the nodes in the layer below (see Fig. \ref{Architecture}).
We recently derived the exact output distribution of such machines
for arbitrary width $k$ and depth $n$ \cite{Fink2026}.
Here, we prove that simplicity bias emerges and strengthens with every additional layer.
At depths of order $2^k$, the output probability approaches
an exponential dependence on Kolmogorov complexity $K$:
	\begin{equation*}
		\Pr(F) \propto 2^{-K(F)}.
	\end{equation*}
This scaling law is in agreement with empirical observations across a broad range of input-output maps, 
including RNA folding, L-systems, self-assembly, gene regulatory networks, deep neural networks and dynamical systems 
\cite{Dingle2018, Johnston2022, Mingard2025, Hamzi2024}.
\\ \indent
Although an individual simple function is exponentially more likely than a complicated one, 
there are exponentially more complicated functions. 
To understand how these effects compete, we ask how likely an output is to have a given complexity.
We prove that every additional layer shifts this distribution towards simpler outputs. 
Remarkably, at depths of order $2^k$, the probability that an output is compressible by a fixed fraction of its length 
is of order one, whereas the corresponding probability for uniformly random outputs is exponentially small.
\\ \indent
Our results suggest that simplicity bias may be a generic consequence of hierarchical composition rather than a byproduct of biological evolution or machine learning. 
Repeated composition of random local rules can systematically favor simple global outcomes, even when the local rules themselves have no preference for simplicity.
This may also help explain a fundamental puzzle in machine learning:
deep neural networks contain vastly more parameters than are needed to fit their training data, 
yet they are resistant to overfitting and generalize well to unseen examples \cite{Zhang2021}. 
Our results suggest that the mapping from local network parameters to global output functions 
could favor simple functions, helping to explain this resistance to overfitting. 
\\ \noindent {\sf\textbf{Input-output map}} \\
A deep-layered machine maps many local Boolean functions to a single global Boolean function.
Each input assigns one of the $2^{2^k}$ possible Boolean functions
to each of the $n k + 1$ network nodes (the gray nodes in Fig.~\ref{Architecture}).
The output is the global Boolean function $F$ of the $k$ arguments (the white nodes in Fig.~\ref{Architecture}),
with $2^{2^k}$ possible outputs.
For example, for $k=2$ and $n=1$, $16^3$ inputs map to 16 outputs,
as illustrated in Fig.~\ref{HeroShot}.
\\ \indent
In our previous paper \cite{Fink2026}, we proved that
the probability of generating an output $F$ depends only on
its Hamming weight $w(F)$, the number of ones in its truth table,
and not on the particular arrangement of those bits.
For convenience, set $\ell=2^k$.
We write $q_i(n)$ for the probability of an output with
Hamming weight $i \in [0, \ell]$,
and $p_j(n)$ for the probability of Boolean function $j \in [1, 2^\ell]$.
Since there are $\binom{\ell}{i}$ functions with Hamming weight $i$,
\begin{equation*}
    \textstyle p_j(n) = q_i(n)\big /\binom{\ell}{i},
    \qquad i=w(F_j).
\end{equation*}
\indent
The Hamming-weight distribution evolves according to
\begin{equation*}
    \mathbf q(n)=\mathbf A^n\mathbf q(0),
\end{equation*}
where $\mathbf A$ is the $(\ell+1)\times(\ell+1)$ transition matrix derived in \cite{Fink2026}.
Writing $x=j/\ell$ and $y=1-j/\ell$, its entries are
	\begin{equation*}
		\textstyle A_{ij}=\binom{\ell}{i}x^i y^{\ell-i},
		\qquad i,j \in [0,\ell].
	\end{equation*}
\indent
Initially, all Boolean functions are equally probable.
As depth $n$ increases, their probabilities change in two ways.
First, the probabilities of the bulk of the outputs (everything but true and false) 
become {\sf U}-shaped when plotted as a function of Hamming weight.
Second, probability steadily transfers from the bulk to the constant outputs true and false.
In the large-depth limit, true and false occur with probability $1/2$ each, while the probabilities of the bulk outputs vanish.
\\ \indent
Since complementary Boolean functions are equally likely, $q_i(n)=q_{\ell-i}(n)$. 
We therefore combine Hamming weights $i$ and $\ell-i$ into a single folded Hamming class $i$.
With $m=\ell/2$, this gives the folded distribution
	\begin{equation*}
		\tilde q_i(n) = (2-\delta_{im})q_i(n),
    		\qquad i \in [0, m].
	\end{equation*}
The factor $2-\delta_{im}$ ensures that the middle Hamming weight is counted only once.
\\ \indent
The $(m+1)\times(m+1)$ folded transition matrix $\tilde{\A}$ is obtained by adding complementary rows of $\A$
and retaining its first $m+1$ columns.
The size of folded class $i$ is the folded binomial coefficient 
	\begin{equation*}
		\textstyle b_i=(2-\delta_{im})\binom{\ell}{i}.
	\end{equation*}
Then we obtain 
	\begin{align}
        		\tilde A_{ij} 	& =  (1 - \delta_{im}/2) \left(A_{ij}+A_{\ell-i,j}\right) 	\qquad i,j \in [0, m]	\nonumber \\
        					& = \textstyle (1 - \delta_{im}/2) \left( \binom{\ell}{i} x^i y^{\ell-i} + \binom{\ell}{\ell-i} x^{\ell-i} y^i \right) \nonumber \\
    					& = b_i (xy)^m \left( (x/y)^{m-i} + ({y}/{x})^{m-i} \right)\!/2 	\nonumber \\
    					& = b_i (xy)^m \cosh \big( (m-i) \log(y/x) \big).
					\label{cosh}
	\end{align}
Remarkably, $\tilde{\A}$ factors into a diagonal matrix, a hyperbolic-cosine matrix, and a second diagonal matrix.	
At $j=0$, the cosh expression is understood by continuity, giving $(1, 0, 0, \ldots)$ for the first column.
The folded distribution then evolves according to
\begin{equation*}
    \tilde{\mathbf q}(n+1)
    =\tilde{\mathbf A}\tilde{\mathbf q}(n).
\end{equation*}
For example, for $k=2$,
	\begin{equation*}
		\tilde{\mathbf A}(2)=
		\begin{scriptsize}
			\frac{1}{4^4} \!\!
			\begingroup \setlength\arraycolsep{1.5pt}
			\begin{pmatrix}
				2 \binom{4}{0} 		& 0				&0	\vspace{5pt} 	\\
				0				& 2 \binom{4}{1}	&0	\vspace{5pt} 	\\
				0				& 0				& \binom{4}{2}
			\end{pmatrix}
			\endgroup
			\begingroup \setlength\arraycolsep{1.5pt}
		\!\! \left(\!
		\begin{array}{ccc}
			\frac{0^0 4^4+4^0 0^4}{2} & \frac{1^0 3^4+3^0 1^4}{2} & \frac{2^0 2^4+2^0 2^4}{2} \vspace{3pt} \\
			\frac{0^1 4^3+4^1 0^3}{2} & \frac{1^1 3^3+3^1 1^3}{2} & \frac{2^1 2^3+2^1 2^3}{2} \vspace{3pt} \\
			\frac{0^2 4^2+4^2 0^2}{2} & \frac{1^2 3^2+3^2 1^2}{2} & \frac{2^2 2^2+2^2 2^2}{2}
		\end{array}
		\!\right)
		\endgroup \!.
	\end{scriptsize}
	\end{equation*}
Like $\A$, the folded matrix $\tilde \A$ is stochastic, so its columns sum to one.
\\ \noindent {\sf\textbf{Simplicity bias}} \\
A Boolean function of $k$ arguments can be represented by its binary truth table:
a string of $2^k$ bits that specifies its value for every combination of its arguments.
For example, for $k = 3$, the function $a b c$ ($a$ {\sc and} $b$ {\sc and} $c$) has truth table 10000000 (adopting the Mathematica convention for ordering), 
since the function is true only when all three of its arguments are true.
\\ \indent
The simplicity of a Boolean function can be measured by the Kolmogorov complexity $K$ of its truth table:
the length of the shortest program that generates the string on a universal Turing machine \cite{Cover2006}.  
Let $\ell = 2^k$.
A Boolean function of $k$ arguments with Hamming weight $w$
can be specified by first giving $w$, with $\ell+1$ possibilities,
then identifying one of the $\binom{\ell}{w}$ functions with that weight.
Since $\log_2(\ell + 1) \simeq  k$,
	\begin{equation*}
		\textstyle
		K(F) \leq \log_2 \binom{\ell}{w(F)} + k + O(1).
	\end{equation*}
\indent
For large $\ell$, this bound is tight up to order $k$ for almost all functions.
To see why, consider $N$ arbitrary binary strings and choose one at random.
The probability of compressing it by at least $s$ bits below $\log_2 N$ is at most $1/2^s$.
Applying this to the $\binom{\ell}{w}$ strings with Hamming weight $w$ gives
$\Pr\left(K(F)\leq\log_2\binom{\ell}{w}-s \right) \leq 2^{-s+O(1)}$.
Setting $s=k$ and taking the complement,
	\begin{equation*}
		\textstyle
		\Pr\left(K(F) > \log_2\binom{\ell}{w} - k \right) \geq 1 - 2^{-k + O(1)}.
	\end{equation*}
Thus, in the limit of large $k$, almost all functions satisfy
	\begin{equation}
		\textstyle
		K(F) = \log_2 \binom{\ell}{w(F)} + O(k)
		\label{InfoContent}
	\end{equation}
We therefore define $K(i)=\log_2\binom{\ell}{i}$ as the effective complexity 
of functions with Hamming weight $i$.
	\begin{figure}[b!]
		\hspace*{-14pt}\includegraphics[width= 1.06\columnwidth]{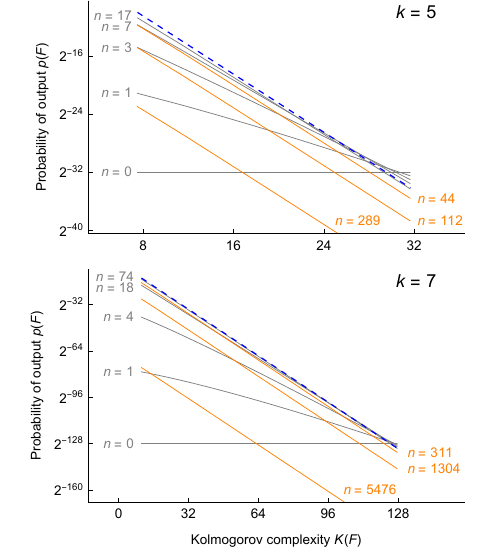}
		\caption{\small
		\textbf{Output probability versus Kolmogorov complexity.}
		The solid curves show the probability of the non-constant outputs.
		As $n$ increases, 
		the curves first rotate clockwise from flat to a line with slope $-1$ (in $\log_2$ space), 
		then fall as the constant outputs true and false (not shown) dominate. 
		The rising curves are gray and the falling curves are orange.
		The dashed blue curve is $P(F) = 2^{-K(F) - k/2}$, predicted by our theory at depth $2^k$;
		the $k/2$ is a finite-$k$ correction
		}
		\label{SimplicityBias}
	\end{figure}
\\ \indent
We now show that simplicity bias exists for every finite $k$ and strengthens with each additional layer.
We then derive its asymptotic form for large $k$.
Both finite-$k$ results follow from
	\begin{equation}
		\frac{\tilde q_{i+1}(n+1)}{\tilde q_i(n+1)} <
 		\frac{\tilde q_{i+1}(n)}{\tilde q_i(n)},
		\qquad i \in [0, m),
    		\label{MA}
	\end{equation}
which follows from the total positivity of order two (TP2) of the folded matrix $\tilde{\mathbf A}$: 
all its $2\times2$ minors are nonnegative. 
We prove TP2 and eq. (\ref{MA}) in the {\it Proof} section below.
\\ \indent
First, since Boolean functions are equally probable at zero depth,
$\tilde q_i(0)=b_i/2^\ell$.
Repeated application of eq.~(\ref{MA}) gives ${\tilde q_{i+1}(n)}/{\tilde q_i(n)} < {b_{i+1}}/{b_i}$.
Thus, for $n > 0$, the probability of an individual function decreases with Hamming class $i$:
	\begin{equation*}
		\tilde q_{i+1}(n)/b_{i+1}<\tilde q_i(n)/b_i.
	\end{equation*}
Since the effective complexity $K$ increases with $i$, we have established simplicity bias for all $k$.
\\ \indent
Second, we define the simplicity-bias exponent $\alpha(i,n)$ as the negative slope of $\log_2 p$ versus $K$:
\begin{equation*}
    \alpha(i,n)=-
    \frac{
        \log_2\left(\tilde q_{i+1}(n)/b_{i+1}\right)
        -\log_2\left(\tilde q_i(n)/b_i\right)
    }{
        K(i+1)-K(i)
    }.
\end{equation*}
Multiplying eq.~(\ref{MA}) by $b_i/b_{i+1}$, taking the logarithm, and dividing by $K(i+1)-K(i)$, we find
	\begin{equation*}
		\alpha(i,n+1)>\alpha(i,n).
	\end{equation*}
Thus, as depth increases, each segment of the $\log_2 p$ versus $K$ curve rotates clockwise, 
as illustrated in Fig.~\ref{SimplicityBias} for $k=5$ and $k=7$.
\\ \indent
But what form does simplicity bias take for large $k$?
From our previous work \cite{Fink2026}, 
we know that the distribution of the bulk of the outputs (everything but true and false) 
rapidly approaches the shape of the leading transient eigenvector of the transition matrix $\A$.
At depths of order $\ell$, the bulk approaches a uniform distribution across the $\ell - 1$ interior Hamming weights, 
while the total bulk probability decays as $\nicefrac{3}{2} \, e^{-n/\ell}$.
Since there are $\binom{\ell}{w}$ functions with Hamming weight $w$, the probability of an individual output $F$ is
	\begin{equation*}
    		p_j(n)\simeq \frac{3e^{-n/\ell}}  {2(\ell-1)\binom{\ell}{w(F_j)}}.	\end{equation*}
Substituting eq. (\ref{InfoContent}) into the above, and noting that $\frac{3}{2(\ell-1)} = 2^{-k+O(1)}$, gives
	\begin{equation*}
		p_j(n)\simeq e^{-n/\ell}2^{-K(F_j)+O(k)}.
	\end{equation*}
Thus random composition produces the characteristic $2^{-K}$ scaling of algorithmic probability, 
despite having no explicit preference for simple functions.
In Fig.~\ref{SimplicityBias}, $2^{-K}$ is indicated by the blue dashed lines.
\\ \noindent {\sf\textbf{Distribution of complexity}} \\
Simplicity bias makes individual simple functions exponentially more probable than complicated ones.
But there is a competing effect: complicated functions vastly outnumber simple ones.
How likely is it that the output of a deep-layered machine is simple?
We show that, as depth increases, these competing effects offset one another, 
making substantially compressed outputs surprisingly common.
\\ \indent
The folded Hamming-weight distribution $\tilde q_i(n)$ gives 
the total probability of obtaining a function in Hamming class $i$.
From eq.~(\ref{MA}), for $b>a$,
	\begin{equation}
		\tilde q_b(n+1)\tilde q_a(n) < \tilde q_b(n)\tilde q_a(n+1).
		\label{NA}
	\end{equation}
Let $\tilde Q_i(n)$ be the corresponding cumulative distribution
	\begin{equation*}
		\Qt_i(n)=\sum_{j=0}^{i}\qt_j(n).
	\end{equation*}
Summing eq. (\ref{NA}) over $a \leq i < b$, we obtain
	\begin{equation*}
		\big(1 - \Qt_i(n+1)\big) \Qt_i(n) < \big(1 - \Qt_i(n)\big) \Qt_i(n+1),
	\end{equation*}
where we have used the normalization of $\tilde q$.
Rearranging,
	\begin{equation}
		\Qt_i(n+1) > \Qt_i(n), \qquad i \in [0,m).
		\label{NM}
	\end{equation}
Now let $C_K(n)$ be the cumulative complexity distribution
	\begin{equation*}
		C_K(n) = \Pr(K(F) \leq K,n).
	\end{equation*}
For almost all functions in class $i$, $K(F)=\log_2\binom{\ell}{i}+O(k)$.
In this approximation, complexity increases with $i$, 
so the ordering in eq. (\ref{NM}) extends to $C_K$:
	\begin{equation*}
		C_K(n+1) > C_K(n).
	\end{equation*}
Thus every additional layer shifts the entire complexity distribution towards simplicity.
\\ \indent
But how dramatic can this shift become?
For large $\ell$, 
	\begin{equation}
		\log_2\binom{\ell}{w} = \ell H_2\left(\frac{w}{\ell}\right) + O(k),
		\label{BinaryEntropy}
	\end{equation}
where $H_2$ is the binary entropy function: $H_2(x)=-x\log_2 x-(1-x)\log_2(1-x)$.
Combining eqs. (\ref{InfoContent}) and (\ref{BinaryEntropy}), we find, for $0 \leq w \leq \ell/2$,
	\begin{equation*}
		\frac{w(F)}{\ell} \simeq H_2^{-1}\left(\frac{K(F)}{\ell}\right).
	\end{equation*}
\indent
At depths of order $\ell$, the bulk (everything but true and false) becomes approximately
uniform across Hamming weights \cite{Fink2026},
with $q_i(n)/P_{\mathrm{bulk}}(n)\simeq 1/(\ell-1)$,
where $P_{\mathrm{bulk}}(n)=1-2q_0(n)$ is the probability
of a nonconstant output.
Let $C_K^{\mathrm{bulk}}(n)$ denote $C_K(n)$ conditioned on nonconstant outputs.
We then obtain
	\begin{equation}
		C_K^{\mathrm{bulk}}(n) \simeq 2H_2^{-1}\left(\frac{K}{\ell}\right).
        		\label{KDist}
	\end{equation}
Thus the exponential preference for simple functions
offsets their exponentially smaller number,
leaving a broad distribution of complexity.
Fig.~\ref{ComplexityDistribution} illustrates this for $k=7$, 
showing how the surviving bulk approaches the asymptotic distribution with $n$.
Meanwhile, an increasing fraction of the total probability accumulates in the constant functions true and false.
\\ \indent
It is well known that the probability of compressing a random binary string by at least $s$ bits is at most $1/2^s$:
    \begin{equation*}
        C_{\ell-s}(0)\leq 2^{-s}.
    \end{equation*}
This exponential suppression is illustrated by the $n=0$ curve in Fig.~\ref{ComplexityDistribution}.
By contrast, eq.~(\ref{KDist}) shows that extensive
compression becomes common among nonconstant outputs,
as shown by the $n = 2^7 = 128$ curve in Fig.~\ref{ComplexityDistribution}.
\\ \noindent {\sf\textbf{\textcolor{black}{Discussion}}} \\ 
Simplicity bias has been empirically observed across a remarkable variety of input-output maps.
Dingle et al. observed that many input-output maps exhibit simplicity bias of the form $P \lesssim 2^{-K}$ \cite{Dingle2018}.
Johnston et al. observed a similar bias in genotype-phenotype maps, arguing that evolution preferentially discovers simple phenotypes because they occupy larger regions of genotype space \cite{Johnston2022}.
Most recently, Mingard et al. observed a $P \sim 2^{-K}$ bias in the output of deep neural networks upon randomly sampling the network parameters \cite{Mingard2025}. 
Table I compares these empirical findings with our result, 
showing that an exactly solvable model reproduces the simplicity bias observed across very different systems.
\\ \indent
As network depth increases, the output distribution passes through three stages (Fig.~\ref{SimplicityBias}).
Initially, all Boolean functions are equally probable.
As $n$ increases, simplicity bias emerges, approaching the characteristic $2^{-K}$ scaling at depths of order $2^k$.
At greater depths, probability accumulates in the two constant functions, each ultimately attaining probability $1/2$.
Throughout this collapse, the relative simplicity bias continues to strengthen.
\\ \indent
Simple functions are exponentially more likely than complicated ones, 
but complicated functions exponentially outnumber simple ones.
Which effect wins?
At large depths, neither.
Each Hamming-weight class contains approximately $2^K$ functions,
each with probability proportional to $2^{-K}$.
Multiplying the two gives approximately equal total probabilities 
across Hamming-weight classes.
Converting from Hamming weight $w$ to complexity $K$ 
through $K\simeq\ell H_2(w/\ell)$ 
then yields the cumulative complexity distribution 
$2H_2^{-1}(K/\ell)$, 
where the factor of two accounts for the symmetry between 
$w$ and $\ell-w$.
Remarkably, repeated composition makes extensive compression substantially probable, 
whereas for random functions it is exponentially unlikely.
\\ \indent
The hierarchical composition of unbiased microscopic update rules 
is sufficient to generate a bias toward simple macroscopic behaviors.
No learning, evolutionary selection, or other form of optimization is required.
How does this happen?
The irreversible nature of the local rules makes previously distinct inputs indistinguishable, 
generating degeneracies that accumulate under repeated composition.
We conjecture that symmetries induced by individual local rules also accumulate under composition, 
constraining the possible outputs and favoring simpler functions.
Simplicity bias may therefore be a manifestation of a more fundamental symmetry bias.
\\ \indent
Our result reveals a surprising connection to Solo\-mon\-off's algorithmic probability \cite{Solomonoff1964}.
In this theory, a universal computer is supplied with random program bits, 
assigning each prefix-free program $p$ a weight $2^{-\text{length}(p)}$.
The algorithmic probability of an output is the sum of these weights over all programs that generate it---which, by the coding theorem, is proportional to $2^{-K}$ \cite{Li2019}.
Thus the probability of an output is set by the length of its shortest description, even though many different programs may generate it.
Starting from a very different perspective, we analytically derive the same exponential decay with Kolmogorov complexity.
This suggests that the characteristic scaling of algorithmic probability may also emerge 
as a statistical consequence of repeated composition of irreversible rules. 
Such a mechanism could explain the simplicity bias observed across learning, evolution, and other many-to-one complex systems.
\vfill              
\onecolumngrid
\begin{center}
\begin{tabularx}{\textwidth}{@{\extracolsep{\fill}}llll}
\emph{Paper} 			& \emph{Systems studied}  							& \emph{Method}	& \emph{Simplicity bias} 	\\
Dingle et al. (2018)		& RNA folding, circadian rhythm, finance model, L-systems  	& Empirical		& $P \lesssim 2^{-K}$ 	\\
Johnston et al. (2022) 	& Self-assembly, RNA folding, gene regulatory network		& Empirical		& $P \lesssim 2^{-K}$	\\
Mingard et al. (2025)	 	& Deep-layered neural networks 						& Empirical		& $P \sim 2^{-K}$		\\
This work 				& Deep-layered machines								& Analytical	 	& $P=2^{-K + O(k)} $ 	
\end{tabularx}
\end{center}
TABLE I: \textbf{Evidence for simplicity bias across input-output maps.}
Previous studies observed an approximately exponential dependence 
of output probability on Kolmogorov complexity $K$. 
Here, we derive this scaling analytically in an exactly solvable model.
\\
\twocolumngrid
\noindent {\sf\textbf{Proof}} \\
We first prove that the folded transition matrix $\At$ is TP2.
To do so, we show that the ratio of two consecutive rows decreases across columns, 
which is equivalent to positivity of the corresponding $2\times2$ minors.
Taking the ratio of consecutive rows using eq.~(\ref{cosh}) and setting $i'=m-i$, we obtain
	\begin{equation*}
		r_j \!=\! \frac{\tilde A_{ij}}{\tilde A_{i+1,j}}
		\!=\! \frac{b_i}{b_{i+1}}
		\frac{\cosh(i'z_j)}{\cosh((i'-1)z_j)},
		\quad
		\textstyle z_j \!=\! \log\frac{\ell-j}{j}.
	\end{equation*}
The logarithmic derivative of $r_j$ with respect to $z_j$ is
	\begin{equation*}
		\frac{d\log r_j}{dz_j}
		= i'\tanh(i'z_j)
		- (i'-1)\tanh((i'-1)z_j)>0,
	\end{equation*}
since $a\tanh(az)$ increases with $a\geq0$ for $z>0$.
The positive derivative shows that $r_j$ increases with $z_j$.
But $z_j$ decreases with $j$, so $r_j$ decreases with $j$,
establishing positivity of the $2 \times 2$ minors.
The first column, $(1,0,\ldots,0)$, also gives
nonnegative minors.
Hence $\tilde{\mathbf A}$ is TP2.
\\ \indent
We prove eq.~(\ref{MA}) by showing
that the first layer establishes an ordering that TP2
preserves at every subsequent layer.
Initially, all Boolean functions are equally probable,
so $\tilde q_i(0)=b_i/2^\ell$. Hence
	\begin{equation*}
		\frac{\tilde q_i(1)}{\tilde q_i(0)}
		= \frac{1}{b_i}\sum_{j=0}^{m}\tilde A_{ij}b_j.
	\end{equation*}
The hyperbolic-cosine form of $\tilde A_{ij}/b_i$
shows that each term decreases with $i$ for $1\leq j<m$
and is constant for $j=m$.
The first column contributes only at $i=0$,
so the ratio decreases with $i$, establishing the base case.
\\ \indent
TP2 matrices preserve monotone likelihood-ratio ordering
\cite{Karlin1968}.
Since the ordering is initially strict and
$\tilde{\mathbf A}$ has positive minors,
successive applications of $\tilde{\mathbf A}$
preserve strictness, giving
	\begin{equation*}
		\frac{\tilde q_i(n+1)}{\tilde q_i(n)} > \frac{\tilde q_{i+1}(n+1)}{\tilde q_{i+1}(n)},
		\qquad i \in [0, m),
	\end{equation*}
for every $n\geq0$.
Rearranging gives eq.~(\ref{MA}).

\end{document}